# Folding Trp-cage to NMR resolution native structure using a coarse-grained model

Feng Ding[†], Sergey V. Buldyrev[*] and Nikolay V. Dokholyan[†]

[†]Department of Biochemistry and Biophysics, The University of North Carolina at Chapel Hill, School of Medicine, Chapel Hill, NC 27599

[*]Center for Polymer Studies, Boston University, Boston, MA 02215

## ABSTRACT

We develop a coarse-grained protein model with a simplified amino acid interaction potential. We perform discrete molecular dynamics folding simulations of a small 20 residue protein – Trp-cage – from a fully extended conformation. We demonstrate the ability of the Trp-cage model to consistently reach conformations within 2Å backbone root-mean-square distance (RMSD) from the corresponding NMR structures. The minimum RMSD of Trp-cage conformations in the simulation can be smaller than 1.00Å. Our findings suggest that, at least for the case of Trp-cage, a detailed all-atom protein model with a physical molecular mechanics force field is not necessary to reach the native state of a protein. Our results also suggest that the success folding Trp-cage in our simulations and in the reported all-atom molecular mechanics simulations studies may be mainly due to the special stabilizing features specific to this miniprotein.

**Keywords:**

Trp-cage, discrete molecular dynamics, protein folding

## Introduction

In 2001 Neidigh et al. discovered that the 18-residue long segment Leu21-Pro38 of exendin-4 – a naturally occur 39 amino acid protein – is the smallest known protein-like folding fragment (1), designated as Trp cage by Barua and Andersen (2). Neidigh et al. (3) have truncated and redesigned the exendin-4 to a 20-residue miniprotein that exhibits cooperative folding transition and is significantly more stable than any other known miniprotein (4-8) ($\Delta G_U \approx 8.6$ kJ mol$^{-1}$ at 3 $^{o}$C). Due to its fast folding kinetics, thermodynamic stability and small size, the Trp cage received a wide attention in computational biophysics community (9-14). These studies have demonstrated the abilities of all-atom molecular mechanics simulations to reach the native state of the Trp cage within approximately 1Å backbone root-mean-square deviation (RMSD) from a completely unfolded conformation.

A central paradigm of molecular biology is that a protein's structure is determined by its amino acid sequence. However, the relationship between a protein's amino acid sequence and its structure (*protein folding problem* (15-21)) remains largely unknown despite a number of important studies (22-35). While a success in the Trp cage folding in computer simulations (10-14) may be perceived as a triumph in solving the protein folding problem, we ask here whether the folding dynamics of the Trp cage is governed by a just few key factors, not specific to the majority of proteins. If captured by physical force-fields employed in molecular mechanics simulations, these factors solely determine the dynamics of the Trp cage and, thus, explain the success of studies in Refs. (9-14).

To answer this question we employ the discrete molecular dynamics (DMD) simulations (36-38). Unlike molecular mechanics simulations driven by physical forces, the DMD simulations are driven by collision events due to ballistic motion of the particles and constraints between these particles (39). Thus, the DMD simulations provide us with an important opportunity to test whether just a set of constraints can be imposed to capture the key factors governing the Trp cage folding dynamics.

The evidence for the key factors determining the Trp cage folding dynamics has been suggested by Neidigh et al. (1), who designed a stable fast folding Trp cage sequence – **NLYIQWLKDGGPSSGRPPPS** – by mutagenesis studies of a common amino acid sequence pattern for Trp cage fold, X**F**XX**W**XXXX**GP**XXXX**PPP**X, where X is any amino acid**.** These three key factors (*i*-*iii*) are listed below. (*i*) Interactions of proline with aromatic residues, such as Pro-Trp, stabilize the Trp cage. Gellman and Woolfson (40) and Neidigh et al. (1) argue that several small proteins, such as WW domains (41), villin headpiece (42), Trp zipper (43), and avian pancreatic polypeptide (44), employ Pro-Trp stacking as a mean of stabilization. (*ii*) The high proportion of proline residue (20%) results in more rigid Trp cage structure than the majority protein structures, drastically reducing the entropy of the Trp cage unfolded state. Gellman and Woolfson (45) pointed out that Trp cage is also rich in Gly residues that contrary to Pro residues increase backbone flexibility and, thus, favor unfolded conformations. We hypothesize that Gly enrichment is essential for the Pro-Trp stacking to occur, and despite their destabilization effect on this protein, Gly residues allow this favorable Pro-Trp interaction. (*iii*) Pitera and Swope (12) also pointed out that a salt bridge between Asp9 and Arg16 in the TC5b variant provides an additional stabilization to the Trp cage.

We develop a coarse-grained protein model, that mimics protein backbone flexibility and side chain packing, and a model of amino acid interactions that are argued to be the key factors determining Trp cage folding dynamics (*i*) – (*iii*). We demonstrate that our model consistently undergoes a folding transition from fully extended conformation to a near-native set of conformations that are within 2.0 Å from the NMR structure (3). We show that some states reach the average NMR structure within less than 1 Å backbone root-mean-square deviation.

## Protein Model

We model protein by beads-on-the-string with beads representing backbone and sidechain heavy atoms. It has been shown that a four-bead DMD model with three backbone beads – N, $C_\alpha$, C' – and one minimalist sidechain bead $C_\beta$ can capture the backbone dynamics of the polypeptides (36, 46, 47). It has long been noticed that the sidechain entropy makes a critical contribution to the protein folding (48). The four-bead model can not be used to estimate the entropy contribution of the sidechains because the $C_\beta$ bead position is solely determined by the backbone dihedral angles, $\Phi$ and $\Psi$. It has also been observed that the packing of different residues in the protein is very important in protein folding and design. The $C_\beta$ beads in the four-bead model can not fully mimic the different sizes for different amino acids. An alternative approach is to position an effective $C_\beta$ bead in the center of mass for the sidechain atoms. Since the correct modeling of backbone dihedral angels requires the van der Waals repulsion between $C_\beta$ atoms and theirs neighboring N, C', and O atoms, the effective $C_\beta$ model can not fully mimic the backbone dynamics of polypeptide. In order to observe protein folding, the model needs to correctly capture not only the backbone entropy but also the sidechain entropy and the size effect for the packing of sidechains. Therefore, to keep the model simple while effectively capture all the important features, we add one or two additional effective sidechain atoms into the four-bead model (36, 46, 47). For the β-branched amino acids – Thr, Ile and Val – we introduce two gamma beads representing the two branches after $C_\beta$. For bulky amino acids – Arg, Lys and Trp – we include an additional $C_\delta$ bead. In Figure 1 we present the schematic diagram of the model protein.

To model hydrogen bonding more accurately, we add the oxygen atoms into the backbone of the original four-bead model (36). For the amino acids that are neither beta-branched nor bulky, the gamma beads are positioned at in the geometrical center of the group of all heavy atoms of the sidechain except $C_\beta$ (see Figure 1). The two effective $C_\gamma$ beads of the β-branched amino acids are centered in the geometrical center of the two groups of heavy atoms forming the branches. For Lys and Arg, the effective $C_\gamma$ beads are located in the position of the actual $C_\delta$ atom. The effective $C_\delta$ bead of Lys is located in position of the charged $N_\zeta$ atom. Similarly, the effective $C_\delta$ bead of Arg plays the role of the positive charge center and coincides with the actual $C_\zeta$ atom. For Trp, the effective $C_\gamma$ bead is centered in the five-atom ring and the effective $C_\delta$ bead is centered in the six-atom benzene ring. We assigned the mass of each bead according to the total mass of the group of atoms it represents.

In order to model the bond lengths and bond angles, we introduce constraints between the nearest and next nearest neighboring beads (38). The parameters are presented in the Table 1. Due to introduction of the gamma and delta beads in the model, we are able to model the sidechain dihedral angles. For proline, we also model the unusual properties of the backbone and side-chain dihedral angels by mimicking the covalent bond between the sidechain and the backbone. We describe in the Methods section the details of the modeling of the sidechain dihedral angles as well as the treatment of the special residue, proline.

**Non-bonded Interactions:** We model amino-acid interactions by assigning square well potentials between pairs of the non-bonded beads (the pairs without bonded constraints). Each bead is modeled as an interacting soft ball with a hardcore radius (*HC*) and its interaction range (*IR*), which are assigned according to statistical analysis of the contacts made between residues (see Methods). We include in our model the hydrophobic interaction $H_{HP}$, salt bridge interaction $H_{SB}$, aromatic interaction between aromatic amino acids $H_{AR}$, aromatic-proline interaction between proline and aromatic residues $H_{AR\text{-}PRO}$, hydrogen bond interaction among main-chains $H_{HB}^{MM}$ , hydrogen bond interactions between side-chains and main-chains $H_{HB}^{SM}$ . Thus the total Hamiltonian of the model, *H*, consists of six contributions:

$$H = H_{HP} + H_{AR} + H_{AR-PRO} + H_{SB} + H_{HB}^{SM} + H_{HB}^{MM} . \quad (1)$$

Here, hydrophobic, salt-bridge, aromatic, and aromatic-proline interactions are solely between the effective sidechain beads of different residues: the beta, gamma, and delta beads. The non-bonded beads of the same residue interact with each other by hard-core repulsion. In order to assign various types of interactions for all pairs of sidechain beads, we categorize all the sidechain beads into six following types (Table 3): hydrophobic (H), amphipathic (A), aromatic (AR), neutral polar (P), positively charged (PC), and negatively charged (NC). One bead can have more than one types, for example, the gamma bead of phenylalanine is both hydrophobic and aromatic (see Table 3).

Only pairwise interactions between sidechain beads are considered in the present model and the potential functions are stepwise:

$$E_{ij} = \begin{cases} +\infty, & d < HC_i + HC_j \\ -2\varepsilon_{ij}/3, & HC_i + HC_j \le d < IR_i + IR_j \\ -\varepsilon_{ij}/3, & IR_i + IR_j \le d < IR_i + IR_j + IR_{ext} \\ 0, & IR_i + IR_j + IR_{ext} \le d \end{cases} \quad (2)$$

where *i* and *j* denote different sidechain beads, the *HC* is the hardcore radius of each bead and the *IR* is the interaction range for each bead (Table 1). The parameter $IR_{ext}$ is introduced to allow an attraction before the two beads come to their interaction ranges. In our study, we set $IR_{ext}$ as 0.75Å. *Hydrophobic* interactions are assigned between two hydrophobic beads or between one hydrophobic and another amphipathic beads if both of the two beads are not aromatic and/or proline. The interaction strengths are assigned as $\varepsilon_{HH}$ and $\varepsilon_{HA}$, respectively. The *aromatic* interactions are assigned between two aromatic beads – namely the $C_\gamma$ of Phe and Tyr, and the $C_\delta$ of Trp – with the strength of $\varepsilon_{AR}$. The *aromatic-proline* interaction is assigned between the $C_\gamma$ bead of proline and the aromatic bead. The interaction strength is $\varepsilon_{AR\text{-}PRO}$. The *salt-bridge* interactions are assigned between the positively charged bead and the negatively charged beads with the strength $\varepsilon_{SB}$. Two beads of the same charge experience the hardcore repulsion.

The hydrogen bond interactions are introduced among the backbones and between the backbone and polar sidechain beads using an algorithm similar to Ref. (36) (see the Methods section for details). The strengths of these interactions are $\varepsilon_{HB}^{MM}$ and $\varepsilon_{HB}^{SM}$, respectively.

In summary, our model has seven interaction parameters: $\varepsilon_{HH}$, $\varepsilon_{HA}$, $\varepsilon_{AR}$, $\varepsilon_{AR\text{-}PRO}$, $\varepsilon_{SB}$, $\varepsilon_{HB}^{MM}$, $\varepsilon_{HB}^{SM}$, and $\varepsilon_\chi$, where $\varepsilon_\chi$ is the interaction strength used to model the dihedral angles (Methods). To fold Trp cage, we have assigned the initial values to the parameters according to available literature (49) and adjust these values using feedback from our folding simulations. In the presented study, we set the parameters of the bonded and non-bonded interaction strengths $\varepsilon_\chi$=1.5ε, $\varepsilon_{HH}$=1.05ε, $\varepsilon_{HA}$ =0.60ε, $\varepsilon_{AR}$=1.80ε, $\varepsilon_{AR\text{-}PRO}$=1.50ε, $\varepsilon_{SB}$=2.70ε, $\varepsilon_{HB}^{MM}$ =5.00ε, and $\varepsilon_{HB}^{SM}$ =2.50ε, where the energy unit, ε, is of the order of 1kcal $mol^{-1}$. Starting from fully extended polymers, we perform molecular dynamics simulations at various temperatures. The temperature unit is related to the energy unit, $\varepsilon/k_B$. The temperature is controlled by Berendsen thermostat (50) with the heat exchange rate equal to 0.1 per time unit. The time unit is the derivative of the units of length, mass and energy, which are defined as Å, atomic mass of carbon $m_C$, and ε, respectively.

## Results and discussions

In order to study the folding process of Trp cage, we perform equilibrium molecular dynamics simulations of a coarse-grained model miniprotein at various temperatures (see Methods) starting from an extend conformation. Throughout this study, the temperature is measured in units of energy ε, divided by Boltzmann constant, $\varepsilon/k_B$ (see Methods). The calculation of RMSD is based on the positions of the backbone $C_\alpha$ atoms and the native state is chosen as the first NMR model of Trp cage (Protein DataBank (51) code: 1L2Y). At very high temperatures, i.e. $T$=1.00, the protein is completely unfolded and remains in the random coil state with the average radius of gyration ($R_g$) of approximately 12 Å. As we decrease the temperature below $T$=0.80, the protein collapses to a compact conformation similarly to the coil-globular transition (52), which is a non-cooperative process and is manifested as the shoulder in the specific heat plot in Figure 2a.

Within the temperature range 0.70<$T$<0.80 the protein remains mostly in the globular state and remains unfolded during most of the simulation time. In Figure 2b, we present a typical simulation trajectories at temperature $T$=0.72. The average radii of gyration of the native, random coil, and fully extended states of Trp cage are approximately 7Å, 12 Å, and 19Å, respectively. The average $R_g$ of the unfolded states at $T$=0.72 is approximately 9.5Å. Thus, the unfolded states in the simulation are significantly collapsed and the extent of reduction of $R_g$ upon folding from these collapsed states is only ≈30%. We also observe that the RMSD of these unfolded states from the native state is on average 4.3Å. Rapid fluctuations in the RMSD suggest that the model protein is mostly present in the unfolded state without populating any specific stable state. According to the studies of Reva et al. (53), the RMSD distribution for a 20-residue protein with randomly selected/constructed globular protein-like structures is Gaussian with the average of 9Å and the standard deviation of 2Å. Therefore, the probability to find a globular structure with RMSD less than 4Å is $10^{-4}$. Thus, the model protein remains in a highly collapsed state with a non-trivial similarity to the native-state, a so-called "molten globular" state (54).

Another important observation during our high temperature simulations is that fluctuations can approach the folded state with RMSD as low as 2Å (Figure 2b), indicating the availability of the native state even at these relatively high temperatures. The native state is not stable at these temperatures and the protein rapidly unfolds to a denatured molten globular state because the potential energy loss due to thermal fluctuation is not sufficient to overcome the gain in the entropic contribution to the free energy which is the product of the temperature and the entropy loss upon folding. By decreasing the temperature, we expect to observe more folded species, defined as the structures with RMSD less than 2Å.

At an intermediate temperature $T$=0.63, we observe the model protein in the folded state with a significantly high probability (Figure 2c). Once the protein reaches the folded state, it remains in the folded state for a long simulation time – longer than $10^4$ time units – and then unfolds. Approximately equal probability of the folded and the unfolded (molten globular) states (Figure 2f) and multiple folding/unfolding transitions along the simulation trajectory (Figure 2c) indicates the proximity of the simulation temperature to the folding transition temperature of Trp cage. To demonstrate the initial folding from the initial stretched-chain conformation, we present in Figure 2d the trajectory of the initial $10^4$ time units. The initial collapse from the stretched chain is very rapid and occurs within 1000 time units as the value of $R_g$ approaches 10Å while the RMSD is still 4Å. After approximately $10^4$ time units, this molten globular state rearranges itself and reaches the folded state with RMSD less than 2Å. In Figure 2e, we present a trajectory for the simulation at low temperature $T$=0.57. At this temperature, the probability to observe the folded state is much larger than that to observe an unfolded state. At low temperatures ($T<T_F$), the folding dynamics becomes slow and the protein model free energy landscape develops kinetic traps upon folding (the first $10^5$ time unit trajectories in Figure 2e). Once the protein folds, it is stable in the folded state with some infrequent and short-lived unfolding fluctuations. In approximately one out of ten simulations at low temperatures, we observe the kinetic trapping which may expand to almost the entire simulation of $5 \times 10^5$ time units (data not shown). However, the potential energy of the traps is always larger than that for the folded state as in Figure 2e.

In Figure 2d, we present the distribution of RMSD for various temperatures. As temperature decreases, the population of folded states increases, so the folding transition temperature can be identified to be approximately $T_F$=0.63. At this temperature, the distribution is bimodal with two peaks of equal area with maxima at 1.7Å and 3.5Å corresponding to folded and unfolded states respectively.

As shown in Figure 2, our simplified model can reproducibly reach the folded state with an average RMSD of less than 2Å and can reach structures with RMSD smaller as 1.0Å in the wide range of temperatures. To characterize the structure of the folded state obtained in DMD simulations, we show in Figure 3a,b a typical DMD configuration with RMSD of 0.96Å from two opposite view points. In these figures, we show coarse-grained representation of the side-chains for different residues. In agreement with NMR structures, the trademark residue of Trp cage Trp6 is closely packed with residues Tyr3, Pro12, Pro18, Pro19, forming the core. We also observe the formation of the salt-bridge between the Asp9 and Arg16. The two helices, α-helix1-8 and $3_{10}$ helix around Ser13, coincide with those in the NMR structures. Keeping in mind that our model includes only a set of key interaction and has coarse-grained backbone and side-chain representations with simplified step-wise interaction potential functions, the proximity of the DMD folded state to the experimental native state is not guaranteed *a priori*.

One important question of a protein model with a set of amino acid interaction parameters is whether the potential energy of the native state corresponds to the ground state, i.e. the lowest energy state of all available structures. In order to address this question for our model with the given simple interaction parameters, we present in Figure 3c,d,e the contour plots of the number of states observed in a simulation trajectory with a given potential energy and RMSD at different temperatures. In general, we observe a significant correlation between the potential energy and RMSD for different temperatures. However, even below folding transition temperature, we still observe some outliers: structures with small RMSD but large potential energies, and structures with large RMSD (≈4.0Å) whose potential energy is close to that of the folded states. Nevertheless, the probability to observe these outliers is very low, of the order of $10^{-5}$ (Figure 3c,d,e). Therefore, the entropy of those states is small and thus the corresponding free-energy is higher than that of the folded states with low RMSD and low potential energy. A similar problem of the existence of the outliers has also been observed in the all-atom molecular mechanics studies (9, 13, 55).

The simplified model combined with fast dynamics algorithm gives us the opportunity to study the folding process in many successful folding events starting from the extended chain. We find that the time needed for folding and also the detailed pathways of folding are extremely heterogeneous for different trajectories at different temperatures. However, an initial collapse is common to all of these folding processes.

For the Trp cage, the initial collapse is mainly due to the aromatic and aromatic-proline interactions. These collapsed structures are non-specific, i.e., have no well-defined structural features. We present in Figure 4a,b two different collapsed structures where the aromatic and/or aromatic-proline contacts are present. Although the salt-bridge interaction is assigned to be the strongest term in the side chain interactions, the salt-bridge between Asp9 and Arg16 is not necessarily present in the collapsed states. To better understand the ensemble property of the collapsed states, we calculate the frequency map (Figure 4e) from the trajectories at $T$=0.72. At this temperature the protein is mainly present in the molten globular states which are flexible and can unfold into completely extended states (see Figure 2b,f). A contact between two residues is defined to exist when any of the interacting side-chain beads are within their interaction ranges. In the frequency map of the collapsed state of Figure 4e, the formation of the short range hydrophobic contacts near the N-terminus have high probability. The probability to observe the salt-bridge between Asp9 and Arg16 is only ≈0.2. The long-range contacts between the poly-proline17-19 and the Trp6 and Tyr3 have also low probability due to the non-specific nature of the collapse state (the contacts within the elliptical circles in Figure 4e).

In order to fold from the collapsed molten globular states into its native state, the protein has to develop the native secondary structure. It is interesting to quantify the propensity of different secondary structures in these collapsed states. Following the method proposed by Rose et al. (56), we calculate the propensity of different secondary structures at $T$=0.72 where the protein remains mostly in the molten globular state (Figure 2f). Since the calculation of secondary structure propensity in Ref. (57) is based only on the backbone dihedral angles, the propensity of strand formation actually measures the propensity to be in extended conformations. The dominant secondary structure is random coil-like except that the polyproline17-19 is extended. Interestingly, the probability to observe helixes for residues 2-9 is significant, ≈10%, indicating a strong helical propensity for first half residues of the Trp cage even in the molten globular state.

It is of great interest to study the folding mechanism from many successful folding transitions observed in our simulations. However, our simulations are done in vacuum, in absence of water. The lack of diffusive friction due to absence of surrounding water might lead to artifacts in folding dynamics in the event sequences and time scales of formation of different secondary and tertiary structures. We argue that although the population of different folding pathways might be different with and without the explicit solvent, the analysis of multiple folding transitions in the absence of solvent might provide us the information of the possible pathways.

According to our simulations, the protein in the collapsed molten globular state must form all the secondary structures including the α-helix, $3_{10}$-helix as well as the salt-bridge which are present in the native fold. This rearrangement process is highly heterogeneous. Typically the formation of the fist α-helix is faster than the formation of $3_{10}$-helix. The preformed salt-bridge behaves as a trap for the formation of the $3_{10}$-helix and needs to break in order for the short helix to form. We also observe in some folding processes a folding pathway similar to what is described in Ref. (13): the pre-formed salt-bridge between Asp9 and Arg16 separates two pre-packed sub-cores of Try3, Trp6, Pro12 and the poly-proline17-19 (Figure 4c); the preformed salt-bridge must break in order for the global folding to occur (Figure 4d).

## Conclusions

We reproduce folding of the miniprotein, a 20-residue long Trp cage, using a simplified protein model. Introducing only key interactions that stabilized the Trp cage, namely the aromatic-proline, salt-bridge, and the hydrogen bond interaction, our coarse-grained model of the mini-protein is able to fold into the native state with average RMSD less than 2Å, while some conformations reach the NMR structure with RMSD less than 1.00Å. Several all-atom molecular dynamics studies for the Trp cage were reported to fold into structures with similar backbone RMSD (10-14). In our DMD model, the protein is simplified into a string of inter-connected beads which are interacting with each other via square-well types of interaction potentials. Therefore, our success to reproductively fold Trp cage into its NMR native state suggests that an atomic detailed protein model with sophisticate force fields is not necessary to fold a protein into its native state, at least in the case of Trp cage.

In addition, we find that once the key stabilizing interactions – the aromatic-proline, salt-bridge, and the hydrogen bond interaction – are emphasized, the resulting folding is not very sensitive to actually assigned values of the parameters (data not shown). This persistent ability of our Trp cage model to fold under the emphasis of the important interactions is due to the special sequence and structural properties specific to Trp cage. For instance, the inclusion of a high level of prolines reduces the available conformation space, as well as increases number the aromatic-proline contacts. The aromatic-proline interaction is commonly observed to stabilize the protein-protein and protein-ligand interactions (58). This might also be one of the reasons for the success of different all-atom molecular mechanics studies of Trp cage using different force fields (10-14). Therefore, we conclude that it might be too early to draw the conclusion about the “correctness” of the modern molecular mechanics force fields from the recent success in the all-atom molecular dynamics folding studies of Trp cage and that additional tests on a large set of proteins are necessary.

An important advantage of the coarse-grained model with simplified interaction potential is the ability to reach an effective time scale of the simulation trajectories several orders of magnitude longer than the traditional all-atom molecular dynamics. We show in this study that our model of the mini-protein is able to undergo multiple folding and unfolding transitions in a single simulation trajectory which is yet to be observed in all-atom molecular mechanics simulations.

In our simulations, we observe a significant correlation between the potential energy and RMSD, i.e. small RMSD states usually correspond to low potential energy states. However, we still observe some outliers or decoy states that have low potential energy but high RMSD. It is possible to train the parameters of the model, which, in our simplified case, include only seven interaction variables, to better satisfy the ground-state criteria by trying various potential-training methods such as minimizing the *Z*-score (59) or perceptron learning (60, 61). More detailed potential energy functions of side-chain interactions may also improve the proximity of the folded state of the model to the experimental native state. However, these methods applied to a single protein do not guarantee the transferability to other proteins (62). In order to improve the predictive power of the current model, one has to design transferable potential energy functions using multiple proteins.

## Methods

**Discrete molecular dynamics**: The discrete molecular dynamics algorithm (DMD) is based on pairwise spherically symmetrical potentials that are discontinuous functions of an interatomic distance (38, 63-65). The earliest molecular dynamics simulations (64) were performed with the discrete algorithm, before the advent of continuous potentials and thus the modern molecular mechanics. In DMD all atoms move with constant velocity unless their come to distances where the stepwise potential function changes. At this moment of time their velocities change instantaneously. This change satisfies the laws of energy, momentum, and angular momentum conservations. When the kinetic energy of the particles is not sufficient to overcome the potential barrier, the atoms undergo a hard core reflection with no potential energy change. At each collision time, only the positions and velocities of the two atoms involved in the collision are updated.

**Sidechain dihedral angles:** Since the model contains up to three effective sidechain beads for the amino acids, we are able to model the sidechain dihedral angles $\chi_1$ and $\chi_2$. It is well known that the rotamers have limited freedom of rotation. We model the behavior of rotamers by introducing effective bonds between the C' and the effective $\gamma_1$ bead for $\chi_1$ and between $C_\alpha$ and the effective $\delta$ bead for $\chi_2$, with the following potential,

$$U_{1,4} = \begin{cases} +\infty, & d < d_{\min}; d > d_{\max} \\ \varepsilon_\chi, & d_{\min} \le d < d_0; d_1 \le d < d_2 \\ 0, & d_0 \le d < d_1; d_2 \le d < d_{\max} \end{cases}, \qquad (3)$$

where $d_{min} < d_0 < d_1 < d_2 < d_{max}$ (Figure 5a,b). As it is demonstrated in the schematic diagram of Figure 5a, the values of $d_0$ and $d_1$ and $d_2$ determine the distribution of correct rotamer angles. We calculate the distributions of distances between the effective gamma bead and backbone C' for different amino acids by sampling over thousands of crystal structures from PDB. For instance, we present in Figure 5c the distribution for valine. The parameters related to the constraints for different residues are listed in Table 2. In Figure 5d, we show the distribution of the $\chi_1$ angles for an unfolded poly-valine peptide from DMD simulations. In our model, the gamma and/or delta beads are coarse-grained atoms and if the gamma and/or delta beads for a certain amino acid are very flexible the corresponding $\chi_1$ and $\chi_2$ angles have no well defined values in the frame of current model.

Therefore, in this model we do not assign any constraints to confine the rotamer angles for the amino acids with flexible effective gamma and/or delta beads: Arg, Glu, Gln, Lys, and Met. Trp residue contains a well-defined $C_\delta$ bead and we introduce a similar constraint between the $C_\alpha$ and the $C_\delta$ bead to model $\chi_2$ (see Table 2).

**Proline:** Proline is a special imino acid because its sidechain is linked by a covalent bond with its backbone amide. Therefore, its distribution of the $\chi_1$ angle differs from such distributions for other amino acids. We assign for proline a covalent bond between gamma bead and its backbone nitrogen bead with an average distance as 1.80Å and the allowed fluctuations of ±0.09Å. Covalently connected to its backbone, proline also has unusual Φ angle distributions (Figure 5e). We introduce a constraint between the prime carbon of previous residue and the beta carbon of proline residue with the distance of 3.63±0.05Å. In Figure 5f, we present the distribution of the dihedral angles of proline from a DMD simulation of poly-proline. The experimental and simulated distributions are in agreement with each other.

**Assignment of the hardcore radii and interaction range:** We model the non-bonded interactions by assigning stepwise potentials between pairs. Each bead is modeled as an interacting soft ball with a hardcore radius and its interaction range. To assess the hardcore radii and interaction ranges for various sidechain beads, we make statistical evaluation of the available crystal structures from PDB. First, we define the existence of a contact between two effective sidechain beads, if any two atoms from the two groups of actual sidechain heavy atoms which the two effective beads represent are within 4.5Å from each other. Next, we calculate the distributions of distances between the two effective sidechain beads that are in contact. From this distribution, we estimate the corresponding hardcore radius, *HC*, and the interaction range, *IR*, which are also listed in Table 1.

**Mainchain hydrogen bond interactions:** The algorithm to model hydrogen bond interaction among mainchain of the protein model in discrete molecular dynamics simulations is similar to that in Ref. (36). The difference between the previous model and the current model is that we introduce the backbone oxygen bead. The hydrogen bond interaction is now between the backbone hydrogen bond donor (HBD), nitrogen $N_i$, and hydrogen bond acceptor (HBA), carbonyl oxygen $O_j$. In order to mimic the angular dependence of the backbone hydrogen bond, we introduce three auxiliary constraints: $N_i – C_j$, $C_{\alpha i} – O_j$, *and* $C_{i\text{-}1} – O_j$, which are presented in Figure 6a as the thin dashed lines. In order to assess the interaction ranges for a hydrogen bond, we calculate the four distances for actual hydrogen bonds by sampling over thousands of native structures from PDB. We define a hydrogen bond in the native structures from PDB by the following criteria: (a) the distance of oxygen and hydrogen are within 2.5Å, (b) the angles of $N_iH_iO_j$ and $C_iO_iH_j$ are larger than 90$^o$. The histograms of the four distances are presented in Figure 6b. The distributions of all the distances are Gaussian-like. Therefore, we define the minimum and maximum interaction distances, $d_{\min}^{HB}$ and $d_{\max}^{HB}$, for each of the related pairs according to their average values and variances (listed in Table 4).

When any one of the four pairs, $N_i – O_j$, $N_i – C_j$. $C_{\alpha i} – O_j$, *or* $C_{i\text{-}1} – O_j$, comes to their corresponding $d_{\max}^{HB}$ distance, the program checks if the rest three pairs whether their $d_{\min}^{HB}$ and $d_{\max}^{HB}$ distances. If all the distances are within their constraints, a hydrogen bond is formed and the potential energy loss is $\varepsilon_{HB}^{MM}$. The corresponding oxygen and nitrogen change their types into their hydrogen bonded types, $N_i$' and $O_j$'. Once they change their types, they can not form any other hydrogen bond unless the existing hydrogen bond breaks. The way for the dissociation of the hydrogen bond is similar. Once any one of the four pairs comes to the distance of $d_{\max}^{HB}$ and the kinetic energy is enough to overcome the loss of kinetics energy $\varepsilon_{HB}^{MM}$, the hydrogen bond breaks and the nitrogen and oxygen return to their original types which are able to form hydrogen bonds again.

**Sidechain and mainchain hydrogen bond interactions**: It has also been pointed out (66-68) that the hydrogen bonds between the polar sidechain and main-chains are important for the starting and ending of α-helices and also for the formation of turns in proteins. We introduce this type of hydrogen bond interactions in our model for those polar residues, namely Thr, Ser, Asn, Asp, Gln, and Glu, which are observed to frequently form this type of hydrogen bonds in the PDB structures (69). There are two types of possible hydrogen bonding interactions between sidechain and main-chains:

**(*i*)** *Sidechain beads as hydrogen bond donor.* We allow the polar sidechain gamma beads of Asn, Asp, Gln, Glu, Ser, and Thr to form hydrogen bonds with the backbone nitrogen. In order to mimic the angular dependence of hydrogen bond, we introduce additional constraints between the gamma bead and the two neighboring beads of the corresponding nitrogen beads – prime carbon and alpha carbon – along the backbone. Since the gamma beads are coarse-grained, we do not introduce any constraints between the backbone nitrogen beads and the neighboring beads of the effective gamma beads.

**(***ii***)** *Sidechain beads as hydrogen bond acceptor.* We also allow the polar sidechain gamma beads of Ser, Thr, Asn, and Gln to form hydrogen bonds with the backbone carbonyl oxygen. The auxiliary constraints are between the neighboring prime carbon and the sidechain gamma beads. Sidechain gamma beads of Asn, Gln, Ser and Thr can be either HBD or HBA. For simplicity, we only allow one type of hydrogen bond to be formed at one time. The parameters for the sidechain and backbone interactions are assigned by analyzing the corresponding hydrogen bonds in the PDB structures and are listed in Table 4.

Once a sidechain gamma bead meets a free backbone nitrogen or oxygen at the hydrogen bonding range $d_{\max}^{HB}$, we check the distances of the corresponding constraints between the gamma beads and the neighboring beads of nitrogen or oxygen: $C_\alpha$ and C' beads near the nitrogen or C' bead near oxygen. If all the constraints are satisfied, potential energy is decreased by $\varepsilon_{HB}^{SM}$ and a temporary bond is assigned for the auxiliary pairs so that the orientation is satisfied during the life-time of hydrogen bond. Both the backbone nitrogen/oxygen and the gamma bead change their types upon the formation of hydrogen bond. Once the hydrogen-bonded gamma beads and its corresponding backbone hydrogen partner, backbone nitrogen or oxygen, come to the distance $d_{\max}^{HB}$ again, the dissociation might happen. If the kinetic energy is enough to overcome the gain of potential energy, the hydrogen bond beaks. Upon the dissociation of the hydrogen bonds, the involved beads change their types back to their original types.

Please note the difference in the treatment of the two types of hydrogen bonds which we introduce in order to simplify the algorithm. A hydrogen bond between two backbone beads may form or dissociate if the oxygen-nitrogen distance or any other distance of the three auxiliary interactions becomes equal to its maximal value $d_{\max}^{HB}$. In contrast, a hydrogen bond between a side-chain bead and a backbone bead may form or dissociate only if the donor-acceptor distance becomes equal to $d_{\max}^{HB}$. In this type of hydrogen bonds, the auxiliary bonds act as permanent bonds with infinitely high potential wells and can form or break only simultaneously with the donor-acceptor bond.

**Acknowledgement:** We thank Charles W. Carter Jr., Sargar Khare, and Brian Kulhman discussions. This work is supported by the UNC Junior Faculty Development IBM Fund Award (to NVD). SVB acknowledges the support from NSF.

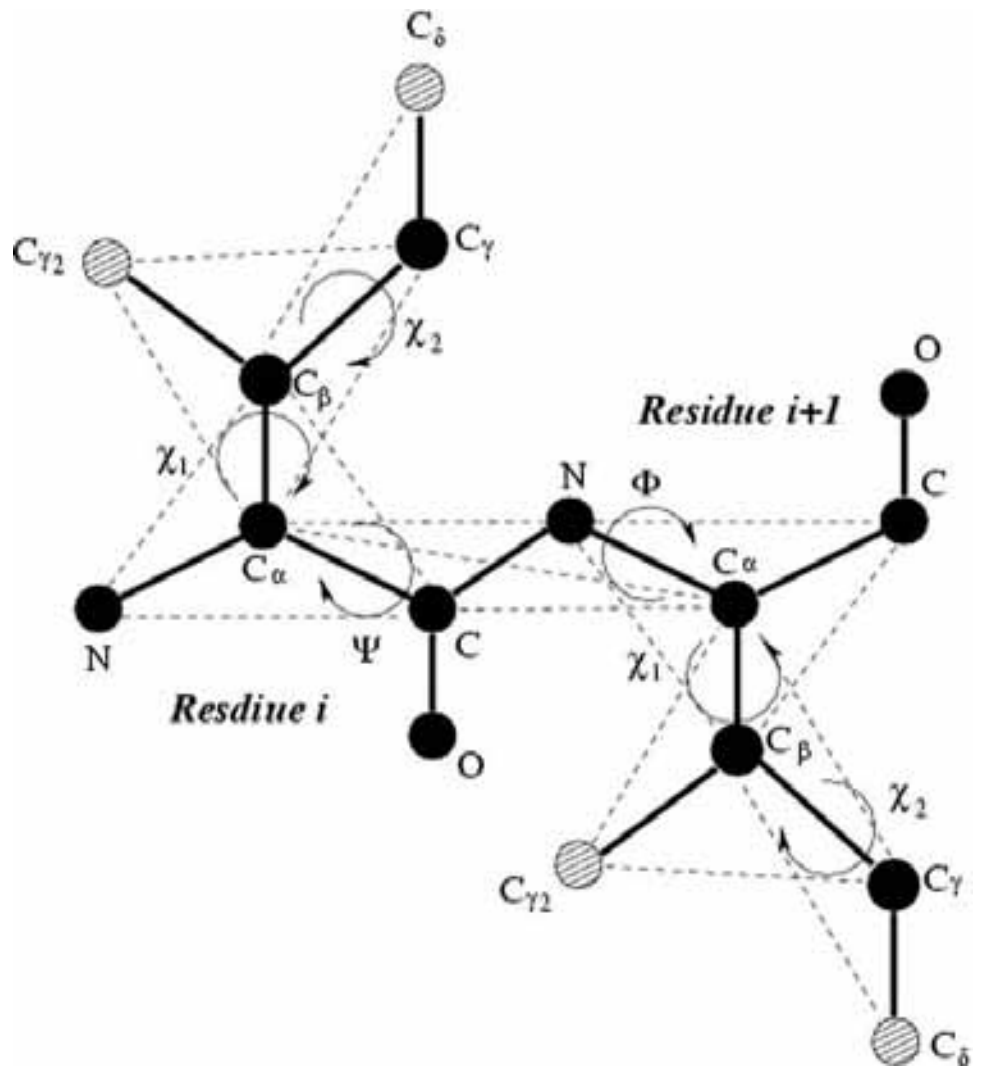

**Figure 1:** The schematic diagram of the model peptide. Only two consecutive residues are presented. The shaded second gamma and delta beads – $C_{\gamma 2}$ and $C_{\delta}$ – indicate that not all amino acids have them. Covalent bonds are represented as thick lines and the constraints that need to fix the bonds angles and the planar property of peptide bonds are denoted as thin dashed lines.

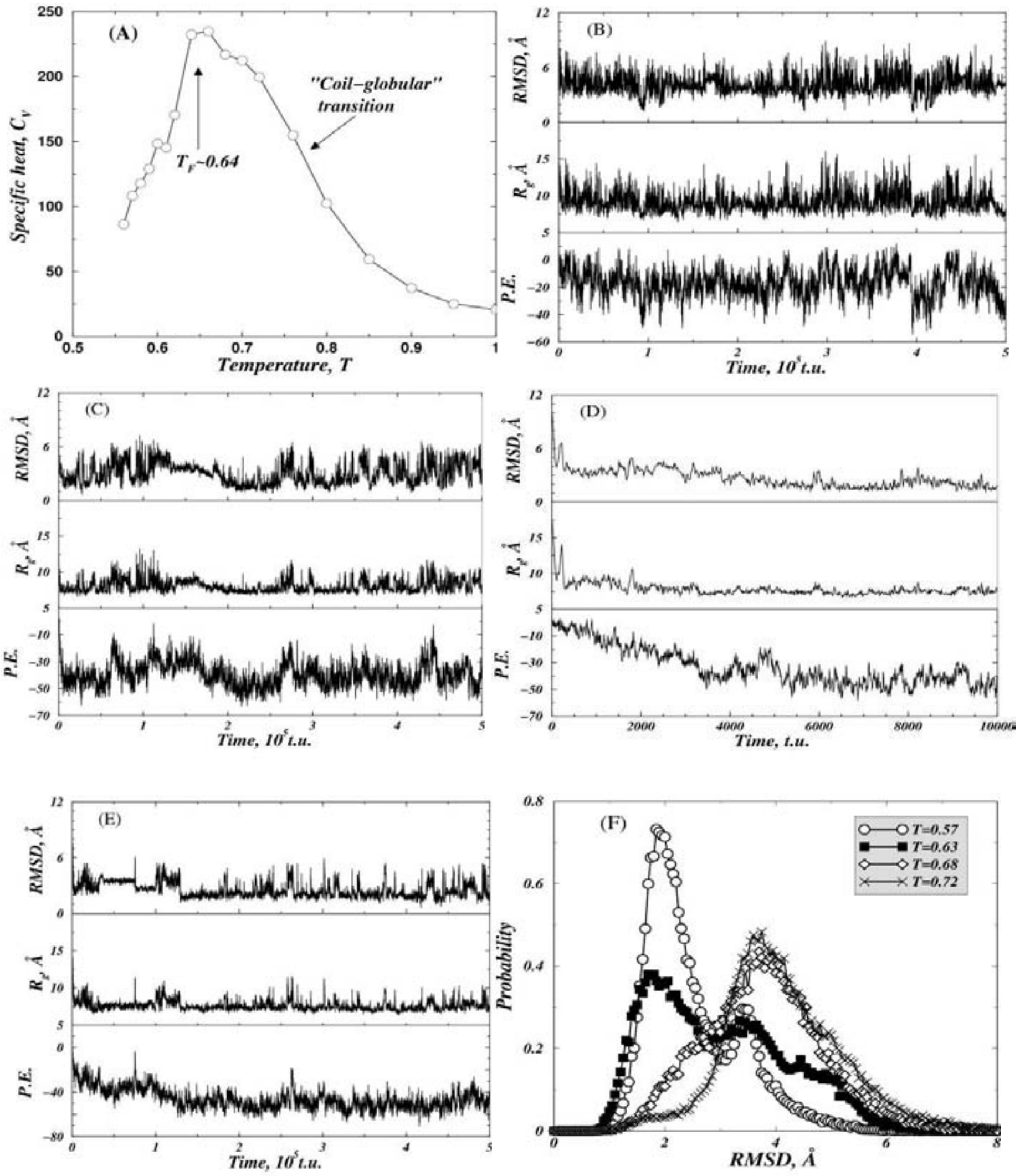

**Figure 2:** The folding thermodynamics of Trp cage. (A) The specific heat $C_v$ as the function of temperature. The potential energy (P.E), radius of gyration ($R_g$) and the $C_\alpha$ RMSD are plotted as the functions of simulation time for different temperatures: (B) T=0.72, (C) T=0.63 and (E) T=0.57. To show the initial collapsing and folding, we present in (D) the folding trajectories of the initial $10^4$ time units at T=0.63. (F) The distributions of the RMSD at different temperatures.

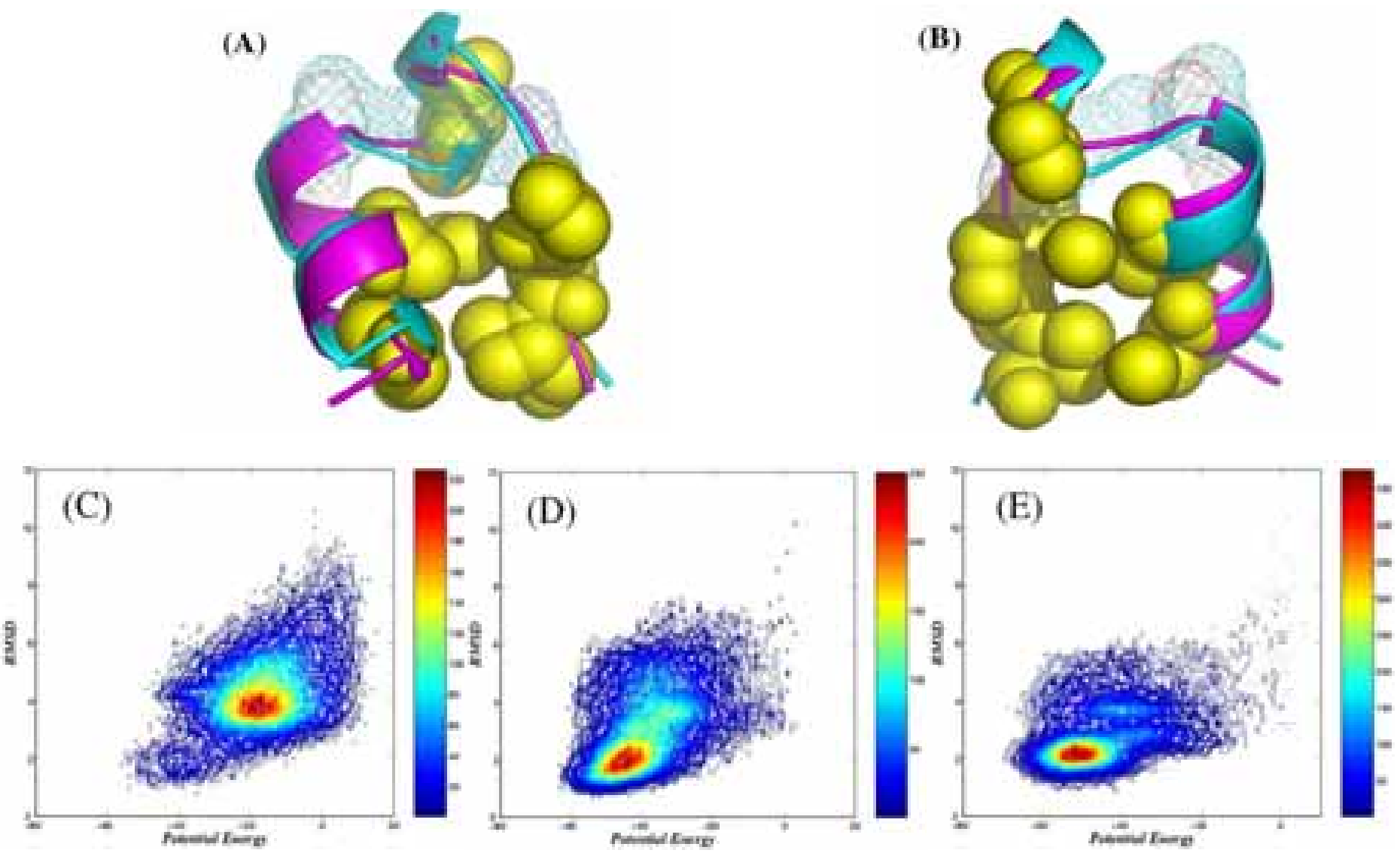

**Figure 3:** The snapshot of one of folded ensemble from DMD simulation is shown in two opposite views (A) and (B). The simulation structure is aligned with respect to the NMR structure which is shown in cartoon representation. The native structure is colored purple and the MD structure is in cyan. In the structure from MD simulations, residues Trp6, Tyr3 and Pro12, 17, 18, and19 are shown in space-filled representation and are colored as golden. We also shown the salt-bridge formed between Asp9 and Arg16 which are drawn as meshed spheres. Since our model is coarse-grained, only the reduced side-chain beads are shown.  The scatter plot of RMSD vs. the potential energy for various temperatures: (C) T=0.72; (D) T=0.63; and (E) T=0.57.

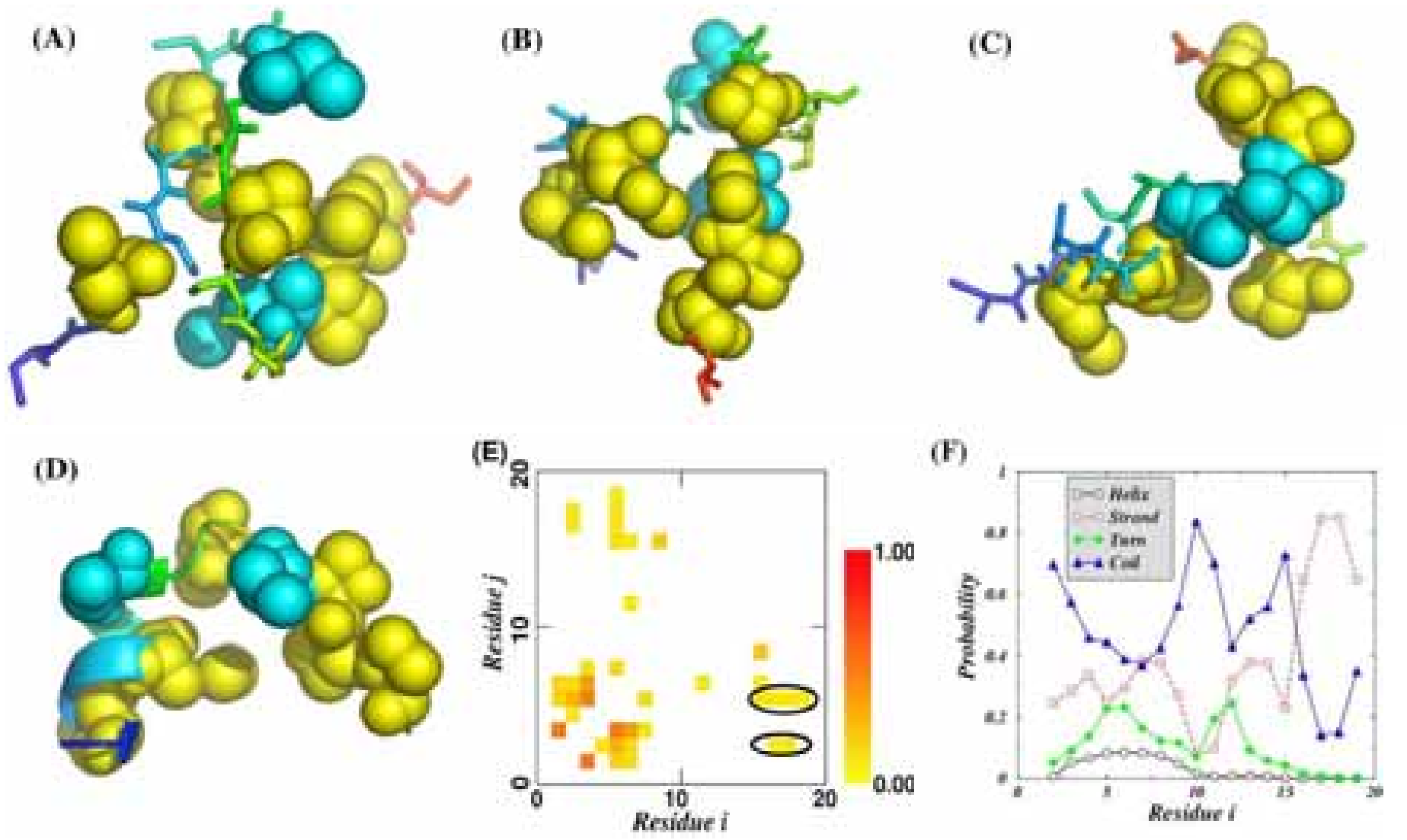

**Figure 4:** (A-B) Two different collapsed "molten-globular" states. (C) A snapshot along the folding pathway is similar to the intermediate observed in Ref. (13). (D) The structure of the model protein that is committed to fold with all the helical secondary structures formed. (E) The contact frequency map of the molten-globular state measured at T=0.72. We only plot the contacts with frequency larger than 0.05. The long range aromatic-proline contacts are encircled by ellipses. (F) The probability of formation various secondary structure elements during simulation at T=0.72.

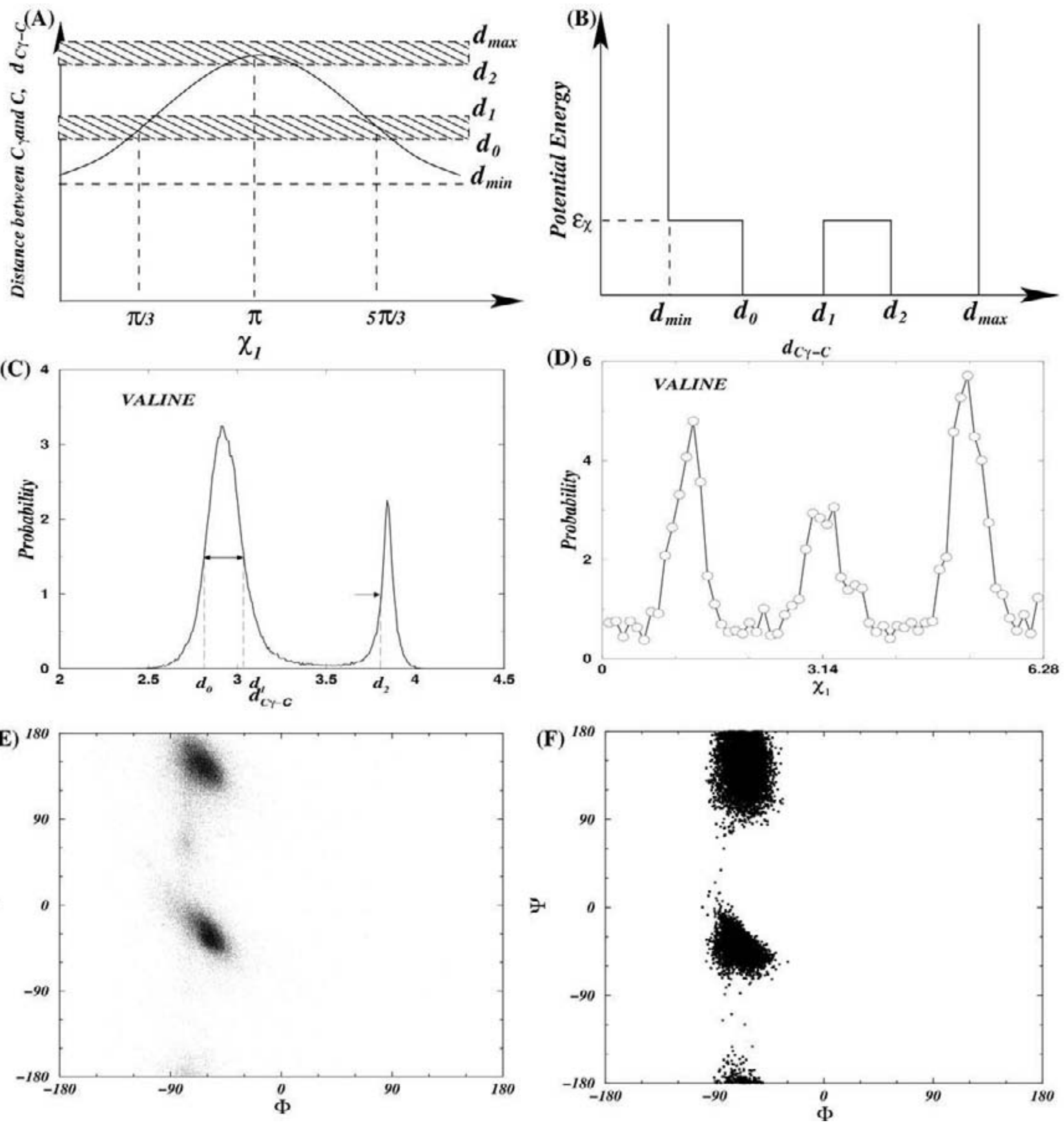

Figure 5: The schematic diagram for the $\chi_1$ constraint. (A) The distance between $C_\gamma$ and C' beads is drawn as the function of rotamer angle $\chi_1$. The shaded regions correspond to the allowed rotamer angle regions around π/3, π, and 5π/3. (B) The introduced potential between the $C_\gamma$ and C' beads. (C) The probability distribution of the distance, $d_{cc'}$ for valine, which is calculated from available PDB structures. (D) The probability of $\chi_1$ angles from DMD simulation of unfolded poly-valine. (E) The Ramachandran plot of proline from (a) various crystal structures from PDB, and from (F) the DMD simulations of a poly-proline peptide.

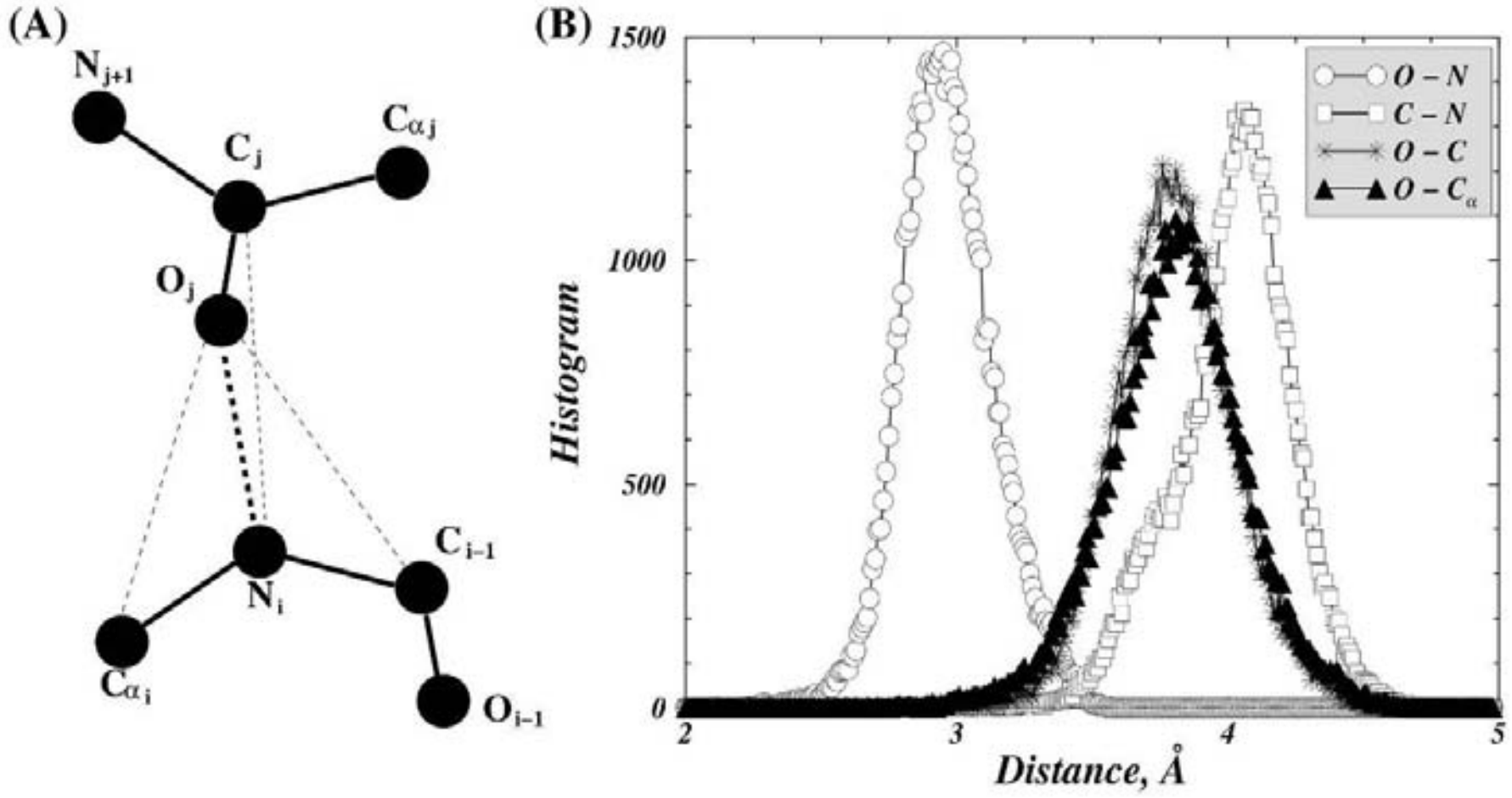

**Figure 6:** (A) The schematic diagram of the hydrogen bond among backbone. Only the backbone beads of the model are shown. The thick dash lines represent the hydrogen bonds and the thin dashed lines indicate the auxiliary constraints for the formation of hydrogen bond. (B) The histogram of distances between the hydrogen bonded oxygen and nitrogen as well as the distance of the auxiliary constraints, which is calculated for the hydrogen bonds in crystal structures.

**Table 1:** We denote the distance of the covalent bonds between beta and gamma beads as $d_{\beta\gamma}$, and the distances of the auxiliary bonds between alpha and gamma beads as $d_{\alpha\gamma}$. For the second gamma beads $C_{\gamma 2}$ of the β-branched residues Thr, Val, Ile, we denote the distance between $C_\beta$ and $C_{\gamma 2}$ beads as $d_{\beta\gamma 2}$, the distance between $C_\alpha$ and $C_{\gamma 2}$ beads as $d_{\alpha\gamma 2}$, the distance between $C_\gamma$ and $C_{\gamma 2}$ beads as $d_{\gamma\gamma 2}$. For the bulky amino acids, Arg, Lys and Trp, we introduce an effective $C_\delta$ bead and denote the distances between $C_\delta$ and $C_\beta$ and between $C_\delta$ and $C_\gamma$ as $d_{\beta\delta}$ and $d_{\gamma\delta}$ respectively. For the distance constraints, we allow a variance of ±2% unless it is specified in the table. All the distances are in the unit of Å and the masses are in the unit of the atomic mass of carbon, $m_C$.

| **Residue** | **$C_\gamma$ Bead** | | | | | **$C_{\gamma 2}$ Bead** | | | | | | **$C_\delta$ Bead** | | | | |
|---|---|---|---|---|---|---|---|---|---|---|---|---|---|---|---|---|
| | $m_\gamma$ | $HC_\gamma$ | $IR_\gamma$ | $d_{\beta\gamma}$ | $d_{\alpha\gamma}$ | $m_{\gamma 2}$ | $HC_{\gamma 2}$ | $IR_{\gamma 2}$ | $d_{\beta\gamma 2}$ | $d_{\alpha\gamma 2}$ | $d_{\gamma\gamma 2}$ | $m_\delta$ | $HC_\delta$ | $IR_\delta$ | $d_{\gamma\delta}$ | $d_{\beta\delta}$ |
| **CYS** | 2.67 | 1.70 | 2.25 | 1.83 | 2.80 | | | | | | | | | | | |
| **MET** | 4.67 | 1.85 | 2.90 | 2.76 | 3.71±0.46 | | | | | | | | | | | |
| **PHE** | 6.00 | 2.00 | 3.20 | 2.91 | 3.79 | | | | | | | | | | | |
| **ILE** | 1.00 | 1.65 | 2.25 | 1.52 | 2.52 | 2.00 | 1.65 | 2.95 | 1.94 | 2.87 | 2.85 | | | | | |
| **LEU** | 3.00 | 2.00 | 3.00 | 1.94 | 3.04 | | | | | | | | | | | |
| **VAL** | 1.00 | 1.65 | 2.20 | 1.52 | 2.50 | 1.00 | 1.65 | 2.20 | 1.52 | 2.50 | 2.49 | | | | | |
| **TRP** | 4.17 | 1.90 | 2.90 | 2.69 | 3.60 | | | | | | | 5.00 | 2.00 | 3.20 | 2.15 | 4.00 |
| **TYR** | 7.33 | 2.00 | 3.20 | 3.30 | 4.16 | | | | | | | | | | | |
| **THR** | 1.00 | 1.65 | 2.25 | 1.52 | 2.49 | 1.33 | 1.35 | 2.20 | 1.43 | 2.41 | 2.42 | | | | | |
| **SER** | 1.33 | 1.35 | 2.25 | 1.45 | 2.43 | | | | | | | | | | | |
| **GLN** | 4.50 | 1.85 | 2.90 | 2.47 | 3.40±0.38 | | | | | | | | | | | |
| **ASN** | 3.50 | 1.75 | 2.70 | 1.94 | 2.88 | | | | | | | | | | | |
| **GLU** | 4.67 | 1.85 | 2.90 | 2.47 | 3.40±0.38 | | | | | | | | | | | |
| **ASP** | 3.67 | 1.75 | 2.70 | 1.94 | 2.88 | | | | | | | | | | | |
| **HIS** | 5.33 | 1.90 | 2.90 | 2.65 | 3.55 | | | | | | | | | | | |
| **ARG** | 3.17 | 1.65 | 2.75 | 2.51 | 3.12±0.38 | | | | | | | 3.33 | 1.85 | 2.90 | 2.47 | 4.30±0.70 |
| **LYS** | 2.00 | 1.65 | 2.75 | 3.40 | 3.12±0.38 | | | | | | | 2.13 | 1.50 | 2.75 | 2.51 | 4.55±0.55 |
| **PRO** | 2.00 | 1.65 | 2.60 | 1.83 | 2.28 | | | | | | | | | | | |

**Table 2:** The parameters of the rotamer constraints: $d_0$, $d_1$ and $d_2$. The parameters $d_{min}$ and $d_{max}$ is not sensitive for sensitive for the correct modeling of the rotamer (Figure 1a). Therefore, we assign 2.0Å and 6.0Å for $d_{min}$ and $d_{max}$, respectively. For Trp the constraint to model $\chi_2$ is between $C_\alpha$ and the $C_\delta$ beads.

| **Residue** | ***$d_0$,Å*** | ***$d_1$,Å*** | ***$d_2$,Å*** |
|---|---|---|---|
| **CYS** | 3.00 | 3.30 | 4.10 |
| **PHE** | 3.70 | 4.18 | 5.12 |
| **ILE** | 2.80 | 3.05 | 3.79 |
| **LEU** | 3.28 | 3.55 | 4.25 |
| **VAL** | 2.80 | 3.05 | 3.79 |
| **TRP** | 3.62 | 4.07 | 4.89 |
| **TYR** | 4.00 | 4.54 | 5.47 |
| **THR** | 2.80 | 3.05 | 3.79 |
| **SER** | 2.68 | 3.06 | 3.68 |
| **ASN** | 3.12 | 3.40 | 4.16 |
| **ASP** | 3.12 | 3.40 | 4.16 |
| **HIS** | 3.57 | 4.05 | 4.87 |
| **TRP($C_\delta$)** | 4.56 | 4.90 | 5.30 |

**Table 3:** The categorization of various sidechain beads. The available types are hydrophobic (H), amphipathic (A), aromatic (AR), neutral polar (P), positively charge (PC), and negatively charged (NC).

| **Residue** | **$C_\beta$** | **$C_\gamma$ bead** | **$C_{\gamma2}$ bead** | **$C_\delta$ bead** |
|---|---|---|---|---|
| **CYS** | **A** | **H** | | |
| **MET** | **A** | **H** | | |
| **PHE** | **A** | **H,AR** | | |
| **ILE** | **A** | **A** | **H** | |
| **LEU** | **A** | **H** | | |
| **VAL** | **H** | **A** | **A** | |
| **TRP** | **A** | **A** | | **H,AR** |
| **TYR** | **A** | **A,AR** | | |
| **ALA** | **A** | | | |
| **GLY** | | | | |
| **THR** | **P** | **A** | **P** | |
| **SER** | **P** | **P** | | |
| **GLN** | **A** | **P** | | |
| **ASN** | **P** | **P** | | |
| **GLU** | **A** | **NC** | | |
| **ASP** | **P** | **NC** | | |
| **HIS** | **P** | **P** | | |
| **ARG** | **A** | **A** | | **PC** |
| **LYS** | **A** | **A** | | **PC** |
| **PRO** | **P** | **A** | | |

**Table 4**: Hydrogen bonding interaction parameters.

| | **Pairs** | $d_{\min}^{HB}, \acute{\AA}$ | $d_{\max}^{HB}, \acute{\AA}$ |
|---|---|---|---|
| **Backbone** | **Ni, $O_j$** | 2.80 | 3.12 |
| | **$N_i$, $C_j$** | 3.80 | 4.23 |
| | **$C_{\alpha i}$, $O_j$** | 3.60 | 4.04 |
| | **$C_{i-1}$, $O_j$** | 3.60 | 4.00 |
| **Thr(HBA)** | **$C_{\gamma 2i}$, $N_j$** | 2.87 | 3.27 |
| | **$C_{\gamma 2i}$, $C_{\alpha j}$** | 3.64 | 4.08 |
| | **$C_{\gamma 2i}$, $C_{j-1}$** | 3.77 | 4.23 |
| **Ser(HBA)** | **$C_{\gamma i}$, $N_j$** | 2.87 | 3.27 |
| | **$C_{\gamma i}$, $C_{\alpha j}$** | 3.64 | 4.08 |
| | **$C_{\gamma i}$, $C_{j-1}$** | 3.77 | 4.23 |
| **ASN(HBA)** | **$C_{\gamma i}$, $N_j$** | 3.52 | 4.04 |
| | **$C_{\gamma i}$, $C_{\alpha j}$** | 4.08 | 4.76 |
| | **$C_{\gamma i}$, $C_{j-1}$** | 4.42 | 4.94 |
| **ASP(HBA)** | **$C_{\gamma i}$, $N_j$** | 3.52 | 4.04 |
| | **$C_{\gamma i}$, $C_{\alpha j}$** | 4.08 | 4.76 |
| | **$C_{\gamma i}$, $C_{j-1}$** | 4.42 | 4.94 |
| **ASN(HBD)** | **$C_{\gamma i}$, $O_j$** | 3.29 | 3.59 |
| | **$C_{\gamma i}$, $C_j$** | 3.16 | 4.00 |
| **GLN(HBD)** | **$C_{\gamma i}$, $O_j$** | 3.50 | 4.06 |
| | **$C_{\gamma i}$, $C_j$** | 4.35 | 4.99 |
| **SER(HBD)** | **$C_{\gamma i}$, $O_j$** | 2.60 | 3.00 |
| | **$C_{\gamma i}$, $C_j$** | 3.53 | 4.13 |
| **THR(HBD)** | **$C_{\gamma 2i}$, $O_j$** | 2.60 | 3.00 |
| | **$C_{\gamma 2i}$, $C_j$** | 3.53 | 4.13 |